\documentclass{ws-p8-50x6-00}
\begin{document}
\title{Nucleon Compton Scattering in Chiral Effective Field Theories}
\author{Thomas R. Hemmert}
\address{Physik Department T39, TU M{\" u}nchen, \\
James-Franck-Stra{\ss}e., D-85747 Garching, Germany \\
Email: themmert@physik.tu-muenchen.de}

\maketitle

\abstracts{
Chiral effective field theories have a long history studying the process of
Compton scattering on the nucleon. In this contribution I want to focus on 
the new developments that have occured since the last Chiral Dynamics
conference in Mainz 1997 \cite{Holstein97}. Moreover, in view of the limited 
time,
I will focus exclusively on the spin-dependent sector, where most of the
recent work has been done.}

\section{Spin Polarizabilities in Real Compton Scattering}

As introduced in the previous talk by Barry Holstein \cite{BRH}, the {\bf 
leading
spin-structure dependent response} of a nucleon in the presence of external
electromagnetic fields can be parameterized via {\bf 4 dipole 
spin-polarizabilities $\gamma_i$},
defined in complete analogy to the more familiar
spin-independent electromagnetic (dipole) polarizabilities
$\bar{\alpha}_E,\,\bar{\beta}_M$.
\begin{eqnarray}
\gamma_{E1}:\quad
        \left(E1\rightarrow E1\right)&\quad\quad &
\gamma_{ M1}:\quad
        \left( M1\rightarrow M1\right)\nonumber\\
\gamma_{E2}:\quad
        \left( M1\rightarrow E2\right)&\quad\quad &
\gamma_{M2}:\quad
        \left(E1\rightarrow M2\right).
\end{eqnarray}
The physics behind these 
spin-polarizabilities thus 
involves excitation of the spin 1/2 nucleon target via
an electric/magnetic dipole transition and a successive de-excitation back
into a spin 1/2 nucleon final state via an electric/magnetic dipole or
quadrupole transition \cite{BGLMN}. As discussed by Holstein
\cite{BRH}, none of these (dipole) 
spin-polarizabilities has been measured directly\footnote{Some information is
known about the particular linear combination
$\gamma_\pi=-\gamma_{M2}-\gamma_{E1}+(\gamma_{E2}+\gamma_{M1})$ from
a dispersion analysis of unpolarized Compton scattering experiments 
\cite{Frank}. For recent efforts to extract the so-called ``forward spin
polarizability'' $\gamma_0=-\gamma_{M2}-\gamma_{E1}-(\gamma_{E2}+\gamma_{M1})$
from measurements of double-polarized pion photoproduction in
the resonance region at MAMI/ELSA and the resulting model dependence see 
ref.\cite{DKT}.} up to know---results of 
future double-polarized Compton scattering experiments on the proton
$\vec{\gamma}\,\vec{p}\rightarrow\gamma^\prime\,p^\prime$ at MAMI, BNL-LEGS
and TUNL, which would suppress the dominant spin-independent physics by 
measuring Compton asymmetries \cite{BGLMN}, are eagerly awaited. In chiral perturbation theory the dipole spin-polarizabilities 
have now been calculated
to next-to-leading order (NLO), 
the results and a comparison with recent dispersion
theoretical calculations are given in Table \ref{dipolespin}.  
The NLO heavy baryon chpt predictions for the essentially unknown dipole spin
polarizabilities of the nucleon can be obtained in closed
form expressions without any free fit parameters. In general the qualitative
agreement between the NLO chiral predictions and the dispersive analyses is
quite good, with the marked exception of the isoscalar spin polarizability
$\gamma_{M2}$, which consistently shows a different sign between the 2
approaches. One also notes that the isovector spin polarizabilities (which
only start at ${\cal O}(p^4)$ in the chiral expansion) tend
to be much smaller than their corresponding isoscalar counterparts, leading to
the non-trivial prediction that the low energy spin structure of proton and
neutron is quite similar~! In column $A$ I have extracted the 
${\cal O}(p^4)$ dipole
spin polarizabilities from the full Compton matrix element by only subtracting
the Born terms as advocated in refs.\cite{spinpolas,Manchester}, whereas in
column $B$ I have subtracted all one-particle reducible graphs from the full
Compton matrix element as advocated in ref.\cite{GHM}. The difference between
these two approaches lies in one particular one-loop diagram, which can be 
seen to primarily affect the large magnetic spin-polarizability $\gamma_{M1}$,
both in the isoscalar and in the isovector channel. Note that in a 
${\cal O}(p^4)$ heavy baryon chpt calculation of the isoscalar $\gamma_{M1}$
spin polarizability the large contribution from an intermediate $\Delta$(1232)
state is still missing, as this effect can only be included at ${\cal
O}(p^5)$ in a full 2-loop calculation \cite{spinpolas}. 
Adding this contribution ``by hand'' as
indicated with the asterisk in Table \ref{dipolespin} shows that procedure $B$
then also leads to a consistent picture for $\gamma_{M1}$, whereas procedure
$A$ requires an unnaturally large higher order cancellation of yet unknown
origin (if the values of the dispersion analyses are to be taken as a serious
benchmark). We note that procedure $B$ has been criticised 
\cite{Judith} in the past few months. In my opionion the only serious criticsm
towards procedure $B$ lies in the fact that the one-to-one correspondence
between a dispersion theoretical ansatz, which only relies on analyticity,
crossing-symmetry etc., and the diagrammatic microscopic chpt calculation gets
lost, because dispersion theory cannot distinguish between one-particle
reducible vs. one-particle irreducible contributions, this clearly being a
microscopic concept having nothing to do with the global properties of
dispersion theory. Further
theoretical investigations are clearly needed to identify the
physics behind the cancellation mechanism needed for procedure $A$. Let me
close this discussion by restating the trivial fact that both procedure $A$ and
procedure $B$ lead to the absolutely identical Compton matrix element---the
differences are only pertaining to a different separation
into a polarizability dependent part and a remainder.

Finally I want to note that a brief summary on recent work regarding the 
extension of the nucleon's spin-dependent response to 
multipolarities beyond the dipole truncation can be found in ref.\cite{TRH}.  
\begin{table}[t]
\caption{Predictions for the isoscalar and isovector dipole
spin-polarizabilities of the nucleon found via 2 different prescriptions 
to ${\cal O}(p^4)$ (i.e. NLO) 
in heavy baryon
chpt, to ${\cal O}(\epsilon^3)$ (i.e. LO) in the small 
scale expansion \protect \cite{SSE}
(SSE$_{\rm{LO}}$) and in dispersion analyses
(DKH,DGPV,BGLM). All results
are given in the units $10^{-4}\,\rm{fm}^4$. 
($*$: $+\,2.5\times10^{-4}\,\rm{fm}^4$ from $\Delta$(1232) pole 
still missing).\label{dipolespin}}
\begin{center}
%\footnotesize
\begin{tabular}{|c||cc|ccc|c|}\hline
$\gamma_{i}^{(N)}$ & chpt$_{\rm{NLO}}^A$ & 
chpt$_{\rm{NLO}}^B$ & DKH\cite{DKH} & 
DGPV\cite{DGPV} & BGLMN\cite{BGLMN} & SSE$_{\rm{LO}}$\cite{spinpolas}  \\
\hline\hline
$\gamma_{E1}^{(s)}$ & $-2.8$ & $-3.0$ & $-5.0$ & $-5.2$ & $-4.5$ & $-5.2$\\
$\gamma_{M1}^{(s)}$ & $+2.8^{*}$ & $+0.4^{*}$ & $+3.4$ & $+3.4$ & $+3.3$ & $+1.4$\\
$\gamma_{E2}^{(s)}$ & $+2.0$ & $+2.0$ & $+2.4$ & $+2.7$ & $+2.4$ & $+1.0$\\
$\gamma_{M2}^{(s)}$ & $+0.3$ & $+0.6$ & $-0.6$ & $-0.5$ & $-0.2$ & $+1.0$\\
\hline
$\gamma_{E1}^{(v)}$ & $+1.4$ & $+1.2$ & $+0.5$ & $+0.8$ & $+1.1$ & - \\
$\gamma_{M1}^{(v)}$ & $+0.5$ & $+0.0$ & $+0.0$ & $-0.5$ & $-0.6$ & - \\
$\gamma_{E2}^{(v)}$ & $-0.2$ & $-0.2$ & $-0.2$ & $-0.5$ & $-0.5$ & - \\
$\gamma_{M2}^{(v)}$ & $-0.1$ & $+0.1$ & $-0.0$ & $+0.5$ & $+0.5$ & - \\
\hline
\end{tabular}
\end{center}
\end{table}

\section{Generalized Spin Polarizabilities}

As discussed in Hyde-Wright's plenary talk \cite{HW}, the pioneering 
Virtual Compton Scattering (VCS) experiment on the proton
$e\,p\rightarrow e^\prime\,p^\prime\,\gamma$ at MAMI is now analyzed 
\cite{nicole}. From the viewpoint of the low energy structure of the nucleon
the only difference to a real Compton scattering experiment is the fact that
the incoming photon is virtual, $Q^2\neq 0$. (Of course one also has to add
the Bethe-Heitler contribution coherently to obtain measurable quantities as
discussed by Hyde-Wright, but here we want to focus on the ``proper'' VCS 
contribution alone.) The Mainz experiment was performed in a special kinematic
regime where one restricts oneself to small energies $\omega^\prime$ of the 
outgoing real photon at fixed three-momentum transfer $|\vec{q}|=600$ MeV 
stemming from the
virtual incoming photon. From a theorist point of view this means that in the 
experiment one wants to be sure, that the de-excitation of the target back
to a spin 1/2 proton via the real photon in the final state 
can be described with
an electromagnetic dipole transition. This is the kinematic
condition that Guichon, Liu and Thomas used for their ground-breaking
definition of generalized
polarizabilities (GPs) \cite{Guichon} and up to now this is the only
theoretical framework we have in order to analyze/discuss low energy VCS
experiments. 

It was shown that in this ``{\bf Guichon limit}'' (i.e. the dipole truncation 
for 
the final state radiation) there is a total of {\bf 6 generalized
(i.e. momentum-transfer dependent) polarizabilities}
$P_{(Y1,Xi)}(\vec{q})$, where $Y1\;(Xi)$ denotes the multipolarity of the
final state (initial state) radiation. In the long-wavelength limit 2 of these 
GPs can be identified with the familiar (spin-independent) electric/magnetic 
polarizabilities:
\begin{eqnarray}
\bar{\alpha}_E & = & - \frac{e^{2}}{4 \pi}
\sqrt{\frac{3}{2}}\,\lim_{\vec{q}\rightarrow 0}
P_{(E1,E1)} (\vec{q}) \,,
\nonumber\\
\bar{\beta}_M & = & - \frac{e^{2}}{4 \pi} \sqrt{\frac{3}{8}}
\,\lim_{\vec{q}\rightarrow 0} P_{(M1,M1)} (\vec{q}) \,.
\end{eqnarray}
(Based on this identification one can also define the momentum-transfer 
dependent generalized dipole polarizabilities
$\bar{\alpha}_E(\vec{q}),\,\bar{\beta}_M(\vec{q})$.) Here we want to focus
on the remaining 4 GPs of the ``Guichon set''
\begin{eqnarray}
P_{(C1,M2)}(\vec{q}),\,
P_{(M1,C2)}(\vec{q}),\,
P_{(M1,C0)}(\vec{q}),\,
\hat{P}_{(C1,(C1,E1))}(\vec{q})\nonumber
\end{eqnarray}
which can be shown to be {\em spin-dependent polarizabilities}.\\
In the long-wavelength limit one can establish a connection to 
two of the dipole spin-polarizabilities of polarized real Compton
scattering\footnote{In the long-wavelength limit one utilizes the connection 
between Coulomb and Electric multipoles.} 
introduced in the previous section:
\begin{eqnarray}
&&\gamma_{M2}\;=\;-\frac{e^2}{4\pi}\,
           \frac{3}{\sqrt{2}}\,\,\lim_{\vec{q}\rightarrow 0}
           P_{(C1,M2)}(\vec{q})\nonumber \\
&&\gamma_{E2}\;=\;-\frac{e^2}{4\pi}\,
                    \frac{3\sqrt{3}}{2\sqrt{2}}\,\lim_{\vec{q}\rightarrow 0}
                    P_{(M1,C2)}(\vec{q})\label{connection}
\end{eqnarray}
The remaining 2 generalized spin polarizabilities $P_{(M1,C0)}(\vec{q}),\,
\hat{P}_{(C1,(C1,E1))}(\vec{q})$ involve longitudinal multipole excitations
$C0,\,C1$ due to the incoming virtual photon and therefore do not have an
analogue in real Compton scattering---they correspond to new
low energy nucleon structure terms which can only be accessed via VCS~!

The ``Guichon set'' of 6 generalized polarizabilities thus has a very intuitive
explanation in terms of multipole excitation/de-excitation, similar to the
multipole basis of the real Compton polarizabilities $\gamma_i$
introduced in the previous
section. Given a theoretical prediction for these 6 GPs, one can calculate the
experimentally accessible response functions
$P_{LL}(\vec{q}),\,P_{LT}(\vec{q}),\,P_{TT}(\vec{q})$ and judge, how well the
theoretical understanding of the microscopic dynamics behind the (generalized)
polarizabilities matches with the real world. For the Mainz experiment the
results are given in Table \ref{VCS}. 
\begin{table}
\caption{\label{VCS} 
Experimental values of the response functions measured at MAMI
at $|\vec{q}|=600$ MeV compared with the
leading order (i.e. ${\cal O}(p^3)$) heavy baryon chpt predictions. 
For comparison
the corresponding values from real Compton scattering which can be obtained in
the long-wavelength limit $|\vec{q}|\rightarrow 0$ are given as well. 
\protect \cite{nicole}.
}
\begin{center}
\begin{tabular}{|ccc|}
\hline
Response& Expt. [GeV$^{-2}$]& Chpt [GeV$^{-2}$] \\
\hline
\hline
$\left[P_{LL}-{1\over \epsilon}P_{TT}\right]_{|\vec{q}|=0}$&$81.0\pm 5.4
\pm 3.3$ &83.5\\
$\left[P_{LT}\right]_{|\vec{q}|=0}$&$-7.0\mp 2.7\mp 1.7$&-4.2\\
\hline
$\left[P_{LL}-{1\over \epsilon}P_{TT}\right]_{|\vec{q}|=600\rm{MeV}}$&$23.7\pm
2.2\pm 0.6\pm4.3$&26.0\\
$\left[P_{LT}\right]_{|\vec{q}|=600\rm{MeV}}$&$-5.0\pm 0.8\pm 1.1\pm 1.4$&
-5.3\\
\hline
\end{tabular}
%\vspace{0.5cm}
\end{center}
\end{table}
One can see that the theoretical values
of the response functions---which use the SU(2) 
${\cal O}(p^3)$ heavy baryon chpt calculation \cite{PRL} of the GPs from 1997 
predating the experiment---are in a good agreement with the experimental
numbers, even though the momentum-transfer in the Mainz experiment is rather
large for such a leading order calculation~! Keeping in mind that the Mainz
experiment only determines one point in the $\vec{q}$-evolution of the response
functions and that certainly further tests (especially at smaller values of 
momentum-transfer) are needed before one can establish a definite picture of
the momentum-dependence of the underlying GPs, I want to discuss the physics 
behind the $\vec{q}$-evolution of the generalized polarizabilities from the 
point of view of chiral effective field theory.

\begin{figure}[t]
\vspace{-1cm}
\centerline{\psfig{file=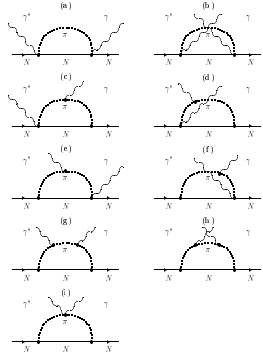,height=8cm}}
\vspace{-2cm}
\caption[diag]{\label{diags} ${\cal O}(p^3)$ $N\pi$-loop diagrams for
Nucleon Compton scattering.}
\end{figure}

For a ${\cal O}(p^3)$ Heavy Baryon calculation of the ``proper'' VCS 
matrix-element---which provides the {\em 
leading order result for the GPs}---one has to evaluate the same 9 one-loop
$\pi N$-diagrams shown in Figure \ref{diags} as for leading order real
Compton scattering with the initial photon now being virtual. To this (leading)
order no new diagrams/counterterms appear beyond the ones also present in real
Compton scattering. It is precisely the simple physics contained in the 
diagrams of 
Figure \ref{diags} evaluated in ref.\cite{PRL} which completely determines the 
$\vec{q}$-dependence of the GPs and results in the theoretical numbers
displayed in Table \ref{VCS} when put into the formulae of the response
functions. The remarkably simple picture behind the physics of the GPs that
emerges suggests that it is the pion-cloud of the nucleon---which can be so 
easily
excited because the chiral symmetry of the underlying QCD-lagrangian is 
spontaneously broken at low energies---that governs the $\vec{q}$-evolution at
small momentum transfer, leading to a typical scaling behavior of 
\begin{eqnarray}
P_{(Y1,Xi)}(\vec{q})\quad\sim\quad\frac{\vec{q}^{\;2}}{m_\pi^2}
\end{eqnarray}
The relevant mass scale thus is the light mass of the (quasi-) Goldstone boson
which {\em in the low momentum regime} makes the effects of the pion-cloud 
dominant over other $\vec{q}$-dependent effects like the 
excitation/de-excitation of nucleon-resonances via transition form factors 
with a typical mass scale of $M_B\sim 1$ GeV. It is obvious that such a regime
should exist for the VCS process due to the (naive) scaling arguments given 
above, however, the positive surprise seems to be that this regime where the
chiral physics dominates over the usual baryon (and vector-meson) resonance
physics seems to extend over a larger range of $|\vec{q}|$ than expected.
By the time of the next Chiral Dynamics workshop we should have a better
understanding of how far in $|\vec{q}|$ the leading order chiral dynamics given
in Figure \ref{diags} provides a sufficient description 
and where the baryon/vector-meson resonance
physics starts taking over. On the experimental side this fall the second VCS 
experiment on the proton
will start taking data at Bates \cite{Jeff} at much lower $\vec{q}$, which
should provide a strong constraint on any theoretical description of the GPs.

The final development in VCS which I want to discuss brings me back to the
spin-sector. Although the pioneering VCS experiment at MAMI was an unpolarized
experiment, it was noted \cite{Baryons}
that the chpt predictions for the spin-dependent generalized polarizabilities
entering the response functions give quite large contributions, even changing
the sign in the case of $P_{LT}(|\vec{q}|=600\rm{MeV})$ for the Mainz
kinematics~! This situation is quite in contrast to the situation in
unpolarized real
Compton scattering, where the spin-polarizabilities tend to be small effects
completely masked by spin-independent physics at low energies. On the 
experimental side this observation has led to a new
proposal \cite{GDH} to perform a second VCS experiment at MAMI, which would 
use
polarized electrons to determine a double-polarization asymmetry via
measuring the average polarization of the recoiling final state proton.
This experiment would give access to a different set of response functions
possibly allowing for enough constraints to separate spin-dependent and
spin-independent GPs. Note that such an experiment also holds the prospect of
determining the essentially unknown real Compton spin polarizabilities
$\gamma_{M2},\,\gamma_{E2}$ via Eq.(\ref{connection}) by measuring
$P_{(C1,M2)}(\vec{q}),\,P_{(M1,C2)}(\vec{q})$ at small values of 
momentum-transfer
and then extrapolating $|\vec{q}|\rightarrow 0$.

On the
theoretical side this prominence of the spin-effects certainly needs to be 
further investigated. In fact, one might worry that this large sensitivity to
spin-effects gives an indication for the breakdown of the ${\cal O}(p^3)$ heavy
baryon calculation---especially if one remembers the tremendous sensitivity 
of some of the real Compton dipole spin-polarizabilities $\gamma_i$ 
on effects connected 
with $\Delta$(1232) intermediate states. There it was argued \cite{spinpolas}
that only a ${\cal
O}(p^5)$ (i.e. 2-loop) heavy baryon calculation (which no group seems to have
on the ``things-to-do'' list until the next Chiral Dynamics workshop) posses
the right operator structure with which one can hope to arrive at a decent 
description
of these important spin-structure quantities. However, nature seems to be very
kind to us in the case of VCS---as long as we stay in the ``Guichon limit''.
First, I want to show you the result of a recent calculation \cite{PRD}
which gives the chiral ${\cal O}(p^3)$ (i.e. leading order) 
momentum dependence of the 4 generalized spin-polarizabilities in closed form
expressions
\begin{eqnarray}
P_{(C1,M2)}^{(3)}(\vec{q})&=& -\frac{g_{A}^2}{24\sqrt{2}\;\pi^2
                              F_{\pi}^2|\vec{q}|^2}\left[1-g({|\vec{q}|\over
                              2m_\pi})
                              \right]\nonumber\\
P_{(M1,C2)}^{(3)}(\vec{q})&=&-\frac{g_{A}^2}{12\sqrt{6}\;\pi^2F_{\pi}^2
                              |\vec{q}|^2}
                              \left[1-g({|\vec{q}|\over
                              2m_\pi})\right]\nonumber\\
P_{(M1,C0)}^{(3)}(\vec{q})&=&\frac{g_{A}^2}{12\sqrt{3}\;\pi^2F_{\pi}^2}
                              \left[2-\left(2+{3|\vec{q}|^2\over 4m_\pi^2}
                              \right)
                              g({|\vec{q}|\over
                              2m_\pi})\right] \nonumber\\
\hat{P}_{(C1,(C1,E1))}^{(3)}(\vec{q})&=&\frac{g_{A}^2}{24\sqrt{6}\;\pi^2
                              F_{\pi}^2|\vec{q}|^2}
                              \left[3-\left(3+{|\vec{q}|^2\over m_\pi^2}\right)
                              g({|\vec{q}|\over
                              2m_\pi})\right] , \label{p3spin}
\end{eqnarray}
with the functional dependence given by
\begin{eqnarray}
g(x)={{\rm sinh}^{-1}(x)\over x\sqrt{1+x^2}}.\nonumber
\end{eqnarray}
One can clearly see that the scale of the $\vec{q}$-variation in the
generalized spin polarizabilities is given by the pion mass $m_\pi$, as
promised. The only other parameters entering the expressions in 
Eq.(\ref{p3spin}) are the axial vector coupling constant $g_A$ measured in
neutron beta-decay and the pion-decay constant $F_\pi$. The leading order 
chiral prediction is therefore completely fixed. What about possible
large corrections
due to $\Delta$(1232) ? Looking at the multipole content of the generalized
spin-polarizabilities in the Guichon limit Eq.(\ref{p3spin}), 
one realizes that 
$\Delta$(1232) pole-contributions can only enter via an interference between
the large $\gamma N\Delta$ $M1$ coupling and the very small $\gamma
N\Delta$ $E2,C2$ quadrupole couplings, triggering some optimism that a
theoretical description of generalized spin polarizabilities {\em in the
Guichon limit} can be reasonable without explicit Delta degrees of freedom.
We note again that once (in future calculations) one goes beyond the Guichon 
kinematical limit and
allows for additional
independent generalized spin polarizabilities which for example involve 
two large
$M1$ transitions to a $\Delta$(1232) in the intermediate state, then one will
encounter the same situation in spin-dependent VCS as in polarized real
Compton scattering \cite{spinpolas}---in heavy baryon chpt 
there would then be no way around a full ${\cal
O}(p^5)$ (i.e. 2-loop) calculation.

\begin{figure}[t]
\vspace{-1cm}
\centerline{\psfig{file=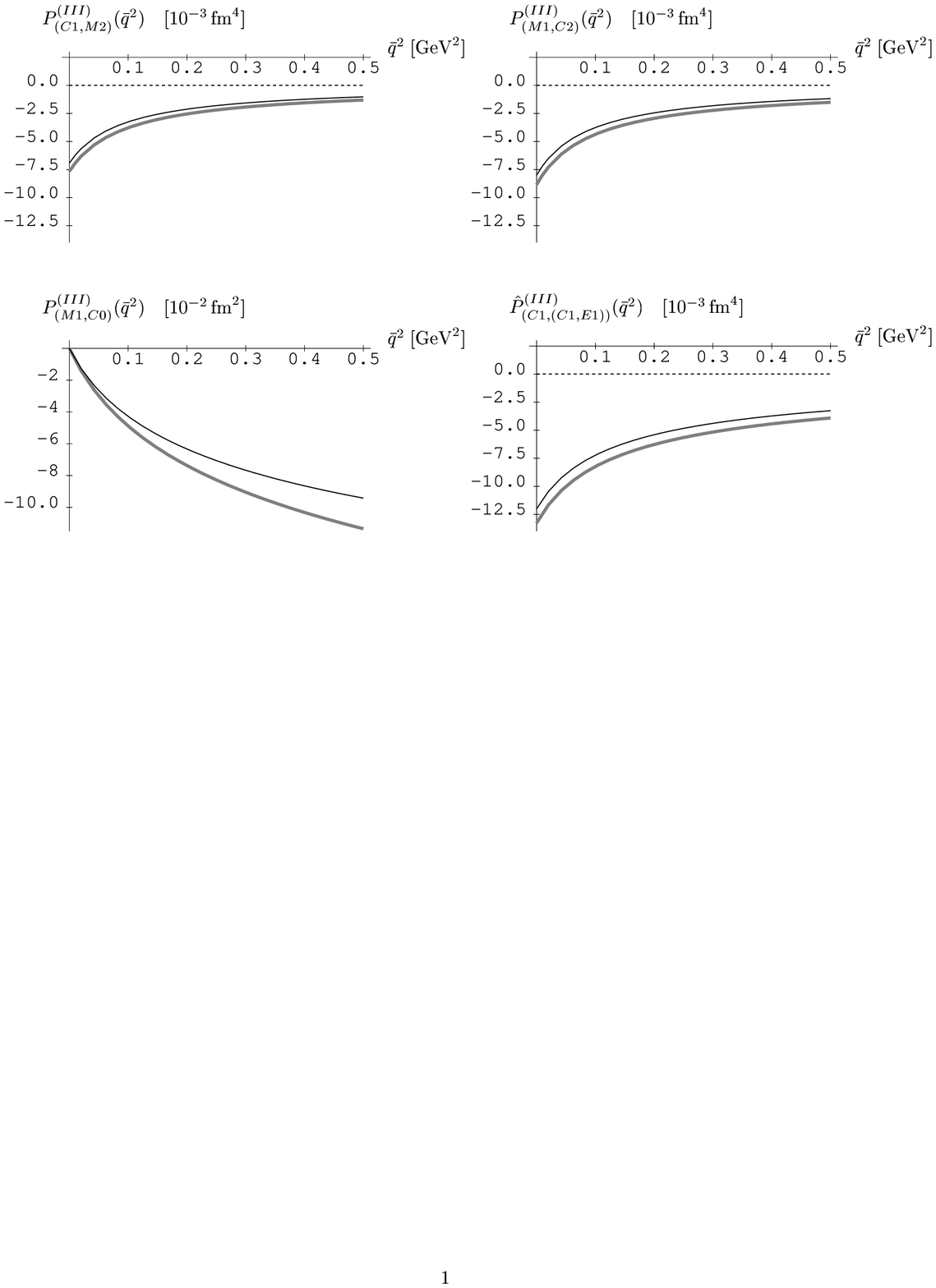,height=14cm}}
\vspace{-8cm}
\caption[diag]{\label{eps3spin} ${\cal O}(\epsilon^3)$ SSE results for
the four generalized spin polarizabilities, 
compared to the ${\cal O}(p^3)$ heavy baryon chpt results of Eq.(\ref{p3spin})
in gray shading \protect \cite{PRD}.}
\end{figure}

Having argued that for the Guichon set of generalized spin-polarizabilities we
do not expect large pole contributions from $\Delta$(1232), what about possible
large corrections due to $\Delta\pi$ intermediate continuum states ? A recent
calculation \cite{PRD} also shows that these contributions are quite small,
completely analogous to the situation of the dipole spin polarizabilities
$\gamma_i$ \cite{spinpolas} in real Compton scattering. Figure \ref{eps3spin}
shows the results of a leading order (i.e. ${\cal O}(\epsilon^3)$) calculation
in the small scale expansion (SSE) \cite{SSE} which contains explicit nucleon, 
pion and delta degrees of freedom. 
As one can clearly see, the heavy baryon
calculation lies quite close to their SSE counterpart, with the largest
discrepancy showing up in $P_{(M1,C0)}(\vec{q})$. The SSE calculation involves
a lot more diagrams but only 2 additional 
parameters---$\Delta=M_\Delta-M_N=292$MeV the location of the 
Delta resonance pole and the strong $\pi N\Delta$ coupling constant $g_{\pi
N\Delta}$, determined consistently within SSE from the width of the Delta
resonance. No other (normalization) adjustment was made, Figure \ref{eps3spin} 
shows the absolute predictions of the 2 quite distinct theoretical 
frameworks, yielding surprisingly similar results. Unfortunately the SSE 
results
cannot be written into a nice analytic form as in Eq.(\ref{p3spin}) but need
a numerical evaluation of the integrals, details can be found in
ref.\cite{PRD}. Certainly further investigations into the generalized spin
polarizabilities are needed, but the new calculations \cite{PRD} suggest, that
chiral effective theories can also provide meaningful
interpretation for the planned polarized VCS experiments focusing on the
spin-dependent response of the nucleon in the presence of external
electromagnetic source-terms. For recent progress utilizing a dispersion
theoretical analysis to predict the $\vec{q}$-dependence of some of the
generalized
spin-polarizabilities starting from pion photo-/electroproduction
data, 
I refer to the talk by M. Vanderhaeghen in the working group \cite{Marc}.   

\section{Summary}

I have reported about new developments since the last Chiral Dynamics workshop
in Mainz in the field of nucleon Compton scattering using chiral effective 
field theories. The spontaneous breaking of chiral symmetry in QCD at low
energies leads to a prominence of $\pi N$ intermediate states which often
dominate
the leading structure dependent response (i.e. the polarizabilities) of the 
nucleon when probed via
external electromagnetic fields. Real and Virtual Compton scattering on the 
nucleon thus provide an excellent laboratory to uncover these signatures of 
chiral
symmetry breaking amidst the usually dominant/overwhelming baryon-resonance
physics. A microscopic understanding of polarizabilities (and their
momentum dependence) in terms of a few simple Feynman diagrams connected to
chiral effective field theories can be given in many cases, leading to a
physical intuition/understanding of the numbers extracted in experiment. 
Several
new experiments will have taken data by the time of the next Chiral Dynamics
meeting improving the data base considerably. The main challenge for the
theoretical efforts is to go beyond the leading order results in a controlled 
and effective approximation.\\
I would like to thank the organizers for giving me the opportunity to present
this overview talk. Many interesting topics had to be skipped (for example the
active field of Compton scattering on the Deuteron \cite{deuteron}) due to lack
of time. See you in J{\" u}lich in 2003~!

%%%%%%%%%%%%%%%%%%%%%%%%%%%%%%%%%%%%%%%%%%%%%%%%%%%%%%%%%

\begin{thebibliography}{99}
\bibitem{Holstein97} B.R.Holstein in Chiral Dynamics: Theory and Experiment;
Proceedings, Mainz, Germany 1997, Eds. A.M. Bernstein, D. Drechsel and W.
Walcher, Springer (Berlin) 1998.
\bibitem{BRH} B.R. Holstein, plenary talk, these proceedings.
\bibitem{BGLMN} D. Babusci et al., Phys Rev. {\bf C58} (1998) 1013.
\bibitem{Frank} F. Wissmann, GDH2000 working group contribution in 
\cite{GDHproc}.
\bibitem{DKT} D. Drechsel, S.S. Kamalov and L. Tiator, preprint no. {\tt
[hep-ph/0008306]} (2000). 
\bibitem{DKH} D. Drechsel, O. Krein, O. Hanstein, Phys.Lett. {\bf B420} 
 (1998) 248.
\bibitem{DGPV} D. Drechsel et al., Phys. Rev. {\bf C61} (2000) 015204. 
\bibitem{SSE} T.R. Hemmert, B.R. Holstein and J. Kambor, J. Phys. {\bf G24}
(1998) 1831; Phys. Lett. {\bf B395} (1997) 89.
\bibitem{spinpolas} T.R. Hemmert et al., Phys. Rev. {\bf D57} (1998) 5746.
\bibitem{GHM} G.C. Gellas, T.R. Hemmert and U.-G. Mei{\ss}ner, 
Phys. Rev. Lett. {\bf 85} (2000) 14.
\bibitem{Manchester} K.B. Vijaya Kumar, J.A. McGovern and M.C. Birse,
Phys. Lett. {\bf B479} (2000) 167. 
\bibitem{Judith} J.A. McGovern, GDH2000 working group contribution in 
\cite{GDHproc}.
\bibitem{TRH} T.R. Hemmert, GDH2000 working group contribution in 
\cite{GDHproc}.
\bibitem{HW} Ch. Hyde-Wright, plenary talk, these proceedings.
\bibitem{nicole} J. Roche et al., Phys. Rev. Lett. {\bf 85} (2000) 708.
\bibitem{Guichon} P.A.M. Guichon, G.Q. Liu and A.W. Thomas, Nucl. Phys. {\bf
A591} (1995) 606.
\bibitem{PRL} T.R. Hemmert et al., Phys. Rev. Lett. {\bf 79} (1997) 22;
Phys. Rev. {\bf D55} (1997) 2630.
\bibitem{Jeff} J. Shaw and R. Miskimen, spokespersons MIT-Bates 
Proposal (1997).
\bibitem{Baryons} N. d'Hose, plenary talk given at Baryons 98; Proceedings of
the 8th International Conference on the Structure of Baryons, Eds. D.W. Menze
and B.Ch. Metsch, World Scientific (1999).
\bibitem{GDH} N. d'Hose, GDH2000 working group contribution in \cite{GDHproc}.
\bibitem{PRD} T.R. Hemmert et al., Phys. Rev. {\bf D62} (2000) 014013.
\bibitem{Marc} M. Vanderhaeghen, working group contribution, these proceedings.
\bibitem{deuteron} c.f. Few Nucleon Processes working group, these proceedings.
\bibitem{GDHproc} Proceedings of the Symposion on
the Gerasimov-Drell-Hearn Sum Rule and the Spin Structure in the Resonance
Region, Mainz (Germany) June 2000; Editor: L. Tiator, World Scientific, 
forthcoming.
\end{thebibliography}
\end{document}